\documentclass{article}

\usepackage[a4paper, total={6in, 8in}]{geometry}

\usepackage{graphicx} 
\usepackage{amsmath} 
\usepackage{amsfonts}
\usepackage{cite}

\usepackage{color}
\usepackage{siunitx}
\usepackage{xcolor}
\usepackage{colortbl}
\usepackage{diagbox}
\usepackage{subcaption}
\usepackage{algorithmic}
\usepackage{algorithm}
\newtheorem{theorem}{Theorem}
\newcommand{\figref}[1]{Fig.\,\ref{#1}}
\newcommand{\Figref}[1]{Figure\,\ref{#1}}
\newcommand{\secref}[1]{Sec.\,\ref{#1}}

\title{Iterative Linear Quadratic Regulator With Variational Equation-Based Discretization}
\author{Katsuya Shigematsu, Hikaru Hoshino\footnote{Corresponding author}, Eiko Furutani}
\date{}

\begin{document}

\maketitle

\begin{abstract}

This paper discusses discretization methods for implementing nonlinear model predictive controllers using  Iterative Linear Quadratic Regulator (ILQR). 
Finite-difference approximations are mostly used to derive a discrete-time state equation from the original continuous-time model.
However, the timestep of the discretization is sometimes restricted to be small to suppress the approximation error.
In this paper, we propose to use the variational equation for deriving linearizations of the discretized system required in ILQR algorithms, which allows accurate computation regardless of the timestep. 
Numerical simulations of the swing-up control of an inverted pendulum demonstrate the effectiveness of this method.
By the relaxing stringent requirement for the size of the timestep, the use of the variational equation can improve control performance by increasing the number of ILQR iterations possible at each timestep in the real-time computation.
\end{abstract}

\section{Introduction}

Model Predictive Control (MPC) is a popular strategy for controlling complex autonomous systems, which solves optimal control problems over a receding horizon and can handle nonlinear models as well as 
constraints \cite{Rawlings2017}. 
It is also used for online trajectory optimization for autonomous robots, where a task-related reference path/trajectory planning and  tracking control of the reference are simultaneously addressed \cite{Neunert2016,Guo2018}.
Various numerical optimization schemes have been developed to overcome the challenge of high computational cost for real-time implementation. 
Among them, direct shooting \cite{Betts2010} is a popular class of methods considered as a computationally efficient approach. 
One of the most famous shooting algorithms is Differential Dynamic Programming (DDP) \cite{Jacobson1970}.
In recent years, one variant of DDP, called Iterative Linear Quadratic Regulator (ILQR) \cite{Li2004}, has attracted great attention especially in the trajectory optimization and machine learning fields (see, e.g., \cite{Levine2014,Grandia2019,Chen2019,Bechtle2020,Lembono2021,Dantec2022}). 
This method has advantages that the size of each sub-problem is time-independent and that the computational complexity grows only linearly with the prediction horizon. 
Nevertheless, applications of MPC has yet been limited by computational limits \cite{Dantec2021}.  

%
In ILQR, like other direct methods, original infinite-dimensional optimal control problems are transcribed into finite-dimensional nonlinear problems, by discretizing the continuous-time problems into discrete-time problems. 
For this, finite-difference methods are mostly used to derive a discrete-time state equation from the original continuous-time model. 
In practice, the forward Euler method is usually employed for the ease of implementation, while higher-order Runge-Kutta methods are also available \cite{Grune2017}. 
The main drawback of these finite-difference methods is that the approximation error increases as the increase in the timestep of the discretization. 
This sometimes restricts the timestep to be too small, since it has to be chosen sufficiently smaller than the time scale of the fastest component of the system. 
The finite-difference methods attempt to approximate the state transition map of the discretized nonlinear state equation, but the algorithm of ILQR only requires the linearization of the state transition map (see \secref{sec:ilqr} for details). 
For this purpose, variational equation \cite{Hirsch2013} can be used to directly calculate the linearization of the state transition map without numerical approximations of the state transition map itself. 
The variational equation is well known in the field of nonlinear science and used for the analysis of nonlinear dynamical systems, but it has hardly been exploited for controller synthesis.

In this paper, we propose an improvement of the algorithm of ILQR by integrating variational equation for discretization. 
By using the variational equation, the timestep of discretization can be made longer than when using the finite-difference methods.
Thus, more time can be used for iterative calculations of ILQR. 
Furthermore, when the prediction interval is fixed, the size of the sub-problems becomes smaller, and computational time can be reduced.
As a result, the possible number of ILQR iteration scales with the square of the timestep, 
and the closed-loop control performance can be significantly improved.  
The effectiveness of the proposed method is demonstrated by numerical simulations of the swing-up control of an inverted pendulum, for which an accurate calculation of the exact optimal trajectory can be performed by using the stable manifold method \cite{Sakamoto2008,Sakamoto2013}. 
A preliminary version of this paper was presented in our conference publication \cite{Shigematsu2023:nolta}. 
This paper adds the following important contributions: while only a simplified algorithm using a quadratic problem for solving subproblems was considered in \cite{Shigematsu2023:nolta}, a complete description of ILQR having the linear time complexity is  considered in this paper; and the benefit of using variational equation is discussed in terms of real-time computation rather than merely the approximation accuracy of the discretization. 

The rest of this paper is organized as follows. 
\secref{sec:ilqr} introduces the basic formulation and algorithm of ILQR, and \secref{sec:ilqr_ve} presents the improved ILQR algorithm using variational equation. 
\secref{sec:simulations} provides numerical simulations of the swing-up control of an inverted pendulum. 
\secref{sec:conclusions} concludes this paper with a summary and
future work.

\section{ILQR Background} \label{sec:ilqr}

ILQR was originally proposed as a trajectory optimization method for computing an open-loop solution of an optimal control problem \cite{Li2004}. 
It can be seen as a variant of the classic Differential Dynamic  Programming (DDP) algorithm \cite{Jacobson1970} and is the control analog of the Gauss-Newton method for nonlinear least squares optimization \cite{Tassa2012}.  
While the original ILQR was not able to deal with constraints or  discontinuous dynamics, recent studies extend ILQR for constrained problems \cite{Howell2019} and hybrid dynamical systems \cite{Kong2023}. 
Below we provide the derivation of a basic unconstrained ILQR algorithm following \cite{Tassa2012}. 
A discrete-time optimal control problem solved by ILQR is presented in \secref{sec:ocp}, the backward recursion used as a main component of ILQR is introduced in \secref{sec:Backwardpass}, and the trajectory optimization method based on ILQR is provided in \secref{sec:ILQR_alg}.

\subsection{Discrete-Time Optimal Control} \label{sec:ocp}

Consider a nonlinear dynamical system with the state $x \in \mathbb{R}^n$, the input $u \in \mathbb{R}^m$, and the dynamics represented as 
\begin{align}
    \dot{x} = f_{\mathrm{c}}(x(t), u(t)) \label{eq:continuous}
\end{align}
where $\dot{x}$ stands for the derivative of $x$ with respect to the time $t$. 
Define a discretization of the continuous dynamics over a timestep $\Delta t$ such that the discrete-time dynamics at the time $t_k := k \Delta t$, for $k = 0,\, 1, \, 2,\, \dots $, are given by 
\begin{equation}
  x_{k+1}= f(x_k,u_k) \label{eq:disstate}
\end{equation}
where 
$x_k := x(t_k)$ and $u_k := u(t_k)$. 
The optimal control problem over the horizon $N$ is given by
\begin{align}
 & \min_U \sum_{i=0}^{N-1}  l(x_i, \, u_i )+l_{\mathrm{f}}(x_N), \label{eq:cost_discrete} \\
 & \mathrm{s.t.} \quad x_0 = x(0), \\
 & \hspace{7mm} x_{i+1} = f(x_i, u_i), \quad \forall i \in \{0, \dots, N-1 \},
\end{align}
where $U := \{u_0, \, u_1 \ldots,\, u_{N-1}\} $ is the control sequence until the horizon $N$ is reached, $l : \mathbb{R}^n \times \mathbb{R}^m \to \mathbb{R}$ represents the running cost, and $l_{\mathrm{f}}: \mathbb{R}^n \to \mathbb{R}$ the terminal cost.
The solution of the optimal control problem is the minimizing control sequence, denoted by $U^\ast(x_0)$, for a particular $x_0$, rather than all possible initial states.

\subsection{Backward Recursion} \label{sec:Backwardpass}

%
%

%
To solve the above problem, DDP/ILQR uses the idea of Bellman recursion to find the optimal control sequence $U^\ast(x_0)$. 
Let $U_i := \{u_i, \, u_{i+1} \ldots ,u_{N-1}\}$ be the sequence of inputs including and after the time $i$, and define the cost-to-go $J_i$ as the partial sum of costs from $i$ to $N$: 
\begin{equation}
  J_i(x_i, U_i) = \sum^{N-1}_{j=i} l(x_j, u_j )+ l_{\mathrm{f}}(x_N)
\end{equation}
with ${x_{i+1}, \, \dots, x_N }$ the sequence of states starting at $x_i$ based on $U_i$ and \eqref{eq:disstate}.
The value function $V(x, i)$ evaluated at the time $i$ is the cost-to-go given the minimizing control sequence: 
\begin{equation}
   V(x, i) := \min_{U_i} J_i(x,U_i). 
\end{equation}
The dynamic programming principle reduces the minimization over an entire sequence of controls to a sequence of minimizations over a single control, proceeding backwards in time:
\begin{equation}
  V(x, i) = \min_{u} [ l(x, u) + V( f(x,u), i+1)], \quad \forall i \in \{ 0, \dots, N-1 \}
  \label{eq:dynamic_programing}
\end{equation}
where the boundary condition is given by $V(x,N) \equiv l_{\mathrm{f}}(x)$. 

Since directly solving the Bellman equation \eqref{eq:dynamic_programing} in real-time is incredibly difficult, DDP/ILQR uses a second order approximation of the argument optimized in the right-hand side of \eqref{eq:dynamic_programing}. 
Specifically, with a nominal control sequence $\bar{U} :=  \{ \bar{u}_0, \, \dots, \bar{u}_{N-1} \}$ and the corresponding nominal trajectory $\bar{X} := \{ \bar{x}_0, \, \dots, \, \bar{x}_N \}$ obtained by applying the control sequence to the dynamical system \eqref{eq:disstate}, define $Q$ to be the argument of the minimum in the right-hand side of \eqref{eq:dynamic_programing} as a function of perturbations around the nominal trajectory: for $i = 0, \dots, N-1$, 
\begin{align}
  Q(\delta x_i, \,\delta u_i, \,i) :=\, & l(\bar{x}_i + \delta x_i, \, \bar{u}_i +\delta u_i, \,i) -l(\bar{x}_i, \bar{u}_i, i) \notag \\
  & +V(f(\bar{x}_i+\delta x_i,\, \bar{u}_i + \delta u_i ), \, i+1) \notag \\ 
  & -V( f(\bar{x}_i, \,\bar{u}_i), \, i+1)
\end{align}
where $\delta x_i := x_i - \bar{x}_i$ and $\delta u_i := u_i - \bar{u}_i$. 
Its second order approximation can be written as 
\begin{align}
 &Q(\delta x_i, \,\delta u_i, \,i) \notag \\
 & \approx 
  \frac{1}{2}
  \begin{bmatrix}
    1 \\
    \delta x_i \\
    \delta u_i
  \end{bmatrix}
  ^\top
  \begin{bmatrix}
    0 && Q_x(i)^\top && Q_u(i)^\top \\
    Q_x(i) && Q_{xx}(i) && Q_{xu}(i) \\
    Q_u(i) && Q_{ux}(i) && Q_{uu}(i)
  \end{bmatrix}
  \begin{bmatrix}
    1 \\
    \delta x_i \\
    \delta u_i
  \end{bmatrix}\label{eq:Q}
\end{align}
with the expansion coefficients given by 
\begin{subequations} \label{eq:expansions}
    \begin{align}
      Q_x(i) := \,& l_x(i) + f_x(i)^\top V_x(i+1), \label{eq:Qx}\\
      Q_u(i) :=\, & l_u(i) + f_u(i)^\top V_x(i+1), \label{eq:Qu}\\
      Q_{xx}(i) :=\, & l_{xx}(i) + f_x(i)^\top V_{xx}(i+1) f_x(i) + V_x(i+1) \cdot f_{xx}(i), \label{eq:Qxx}\\
      Q_{uu}(i) := \,& l_{uu}(i) + f_u(i)^\top V_{xx}(i+1) f_u(i) + V_x(i+1) \cdot f_{uu}(i),  \label{eq:Quu}\\
      Q_{ux}(i) := \, & l_{ux}(i) + {f_u(i)}^\top V_{xx}(i+1) f_x(i) + V_x(i+1) \cdot f_{ux}(i), \label{eq:Qux}
    \end{align}
\end{subequations}
where subscripted variables represent derivatives of the function with respect to the variable, and the index $i$ is used to represent that the derivatives are evaluated at the $i$-th pair $(\bar{x}_i, \bar{u}_i)$, e.g., 
\begin{subequations} \label{eq:linearization}
\begin{align}
  & l_x(i) := \dfrac{\partial l(x,\, u)}{\partial x}\bigg|_{x=\bar{x}_i, u=\bar{u}_i}, \\
  & f_x (i) := \dfrac{\partial f(x,\, u)}{\partial x}\bigg|_{x=\bar{x}_i, u=\bar{u}_i},  \\
  & V_x (i+1) := \dfrac{\partial V(x, \, i+1)}{\partial x}\bigg|_{x=\bar{x}_{i+1} = f(\bar{x}_i,\, \bar{u}_i)}.  
\end{align}  
\end{subequations}
The last terms in \eqref{eq:Qxx}, \eqref{eq:Quu} and \eqref{eq:Qux}, where the dot operation indicates tensor contraction, are used in DDP but ignored in ILQR. 
Since the step computed by DDP is nearly identical to the full Newton step of nonlinear least squares, ILQR can be seen to correspond to the Gauss-Newton Hessian approximation \cite{Tassa2012}.
With this approximation, the second derivative terms with respect to the dynamics ($f_{xx}$, $f_{uu}$, and $f_{xu}$) are neglected, and only the linearized model of the dynamics ($f_x$ and $f_u$) are used.

With this value function expansion, the optimal control input, $\delta u^\ast_i$ can be found by minimizing \eqref{eq:Q} with respect to $\delta u_i$, and we have 
\begin{align}
  \delta u_i^\ast = & \underset{\delta u_i} {\operatorname{argmin}} Q(\delta x_i,\delta u_i, i) \notag \\
  = & -Q_{uu}(i)^{-1}(Q_u(i) + Q_{ux}(i) \delta x_i).
\end{align}
This optimal control input can be split into a feedforward term $d(i)$ and a feedback term $K(i) \delta x_i$ with 
\begin{subequations} \label{eq:opt_control}
    \begin{align}
     d(i) := & - Q_{uu}(i)^{-1} Q_u(i), \\
     K(i) := & - Q_{uu}(i)^{-1} Q_{ux}(i). 
    \end{align}    
\end{subequations}
Plugging the optimal controller into \eqref{eq:Q}, we now have 
\begin{subequations} \label{eq:value_derivatives}
  \begin{align}
V_x(i) & = Q_x(i) - Q_u(i) Q_{uu}(i)^{-1} Q_{ux}(i), \\
V_{xx}(i) &= Q_{xx}(i) - Q_{xu}^\top Q_{uu}(i)^{-1} Q_{ux}(i). 
  \end{align}
\end{subequations}
Now the expansion coefficients in \eqref{eq:expansions} at the time  $i$ can be expressed as a function of the coefficients at the time  $i+1$. 
Thus, the optimal control modification $\delta u_i^\ast$ (or $d(i)$ and $K(i)$) can be computed by recursively computing the coefficients in \eqref{eq:expansions} with the boundary condition given by
\begin{subequations} \label{eq:boundary condition}
\begin{align}
    V_x(N) & = \frac{\partial V(x, N)}{\partial x}\bigg|_{x=\bar{x}_N} = \frac{\partial l_f(x)}{\partial x}\bigg{|}_{x=\bar{x}_N},  \\
    V_{xx}(N) & = \frac{\partial^2 V(x, N)}{\partial x^2}\bigg |_{x=\bar{x}_N}=\frac{\partial^2 l_f(x)}{\partial x^2}\bigg{|}_{x=\bar{x}_N}. 
\end{align}
\end{subequations}
The above process is called the backward pass and used for obtaining the optimal control sequence for the linearized dynamics ($f_x$ and $f_u$) around a given nominal trajectory $(\bar{X}, \bar{U})$ in the ILQR algorithm, where solving the LQR problem for the linearized dynamics and the update of the nominal trajectory is iteratively repeated until convergence.

\subsection{Trajectory Optimization Algorithm} \label{sec:ILQR_alg}

Once the backward pass is completed, a forward pass can be run to compute a new nominal control sequence $\{ \hat{u}_i\}_{i=0}^{N-1}$ and trajectory $\{ \hat{x}_i\}_{i=0}^N$:
\begin{subequations} \label{eq:forward_pass}
  \begin{align}
    \hat{x}_0 & = x(0), \\
    \hat{u}_i & = \bar{u}_i +K(i)( \hat{x}_i - \bar{x}_i ) + \alpha d(i),  \\
    \hat{x}_{i+1} & = f( \hat{x}_i, \, \hat{u}_i).  \label{eq:f} 
  \end{align}
\end{subequations}
where $\alpha \in (0, 1]$ is a line-search parameter detailed below.  
In ILQR, the backwards and forwards passes are run until convergence. 
As discussed in \cite{Tassa2012},     
convergence issues may occur when $Q_{uu}$ is not positive-definite or when the second-order approximation of $Q$ are inaccurate. 
The line-search parameter $\alpha$ introduced above is used to solve the issue that for a general nonlinear system, the new trajectory strays too far from the linearized model's region of validity, and the cost
may not decrease. 
Regularization of $Q_{uu}$ is also added to address the above issues. 
A standard one is to add a diagonal term to the local control-cost Hessian \cite{Liao1991}:
\begin{align}
 \tilde{Q}_{uu} = Q_{uu} + \mu I_m,
\end{align}
where $I_m$ is the $m$-dimensional identity matrix, and $\mu$ plays the role of a Levenberg-Marquardt
parameter, and this modification amounts to adding a quadratic cost around the current control-sequence, making the steps more conservative.  
While there can be several different implementations of line search and regularization to improve the convergence properties of the algorithm, this paper uses a simple implementation as shown in Algorithm\,\ref{alg:ilqr}, where the line search is done by minimizing the actual cost with respect to $\alpha \in \{0.0625, \ 0.125, \,0.25, \,0.5, \,1 \}$, and no regularization term is added. 
See, e.g., \cite{Tassa2012} for improved line search and regularization.


\begin{figure}[!t]
  \begin{algorithm}[H]
      \caption{Trajectory Optimization Using ILQR}
      \label{alg:ilqr}
      \begin{algorithmic}[1]  
      \STATE Initialize $x_0$ and $\bar{U}$
      \STATE $\bar{X} \leftarrow $ Simulate from $x_0$ using $\bar{U}$ and \eqref{eq:disstate}
      \FOR{ $j = 1: 1: N_{\mathrm{iter}} $}
      \STATE /* Linearization */
      \FOR {$i = 0:1:N$}
      \STATE Calculate linearized model $f_x(i)$, $f_u(i)$ and cost derivatives $l_x(i)$, $l_u(i)$, $l_{xx}(i)$, $l_{uu}(i)$, $l_{ux}(i)$ as in \eqref{eq:linearization}
      \ENDFOR
      \STATE /* Backward pass */
      \STATE {Calculate} $V_x(N)$,$V_{xx}(N)$ by \eqref{eq:boundary condition}
      \FOR {$i = N-1:-1:0$}
        \STATE Calculate coefficients $Q_x(i)$, $Q_u(i)$, $Q_{xx}(i)$, $Q_{uu}(i)$, and $Q_{ux}(i)$ from coefficients at $i+1$ using \eqref{eq:expansions} and \eqref{eq:value_derivatives} 
        \STATE Calculate $d(i)$ and $K(i)$ using \eqref{eq:opt_control}
      \ENDFOR
      \STATE /* Forward pass */ 
      \FOR {$i = 0:1:N-1$}
        \STATE Calculate $\hat{X}$ and $\hat{U}$ using \eqref{eq:forward_pass} for each $\alpha$ 
      \ENDFOR
      \STATE $\bar{X},\, \bar{U} \gets \hat{X}$, $\hat{U}$ that minimize the cost function
      \ENDFOR
      \end{algorithmic}
  \end{algorithm}
  \end{figure}

\section{ILQR Using Variational Equation} \label{sec:ilqr_ve}

In this section, we discuss discretization of the continuous-time dynamics \eqref{eq:continuous} to obtain linearized representations of the discrete-time dynamics \eqref{eq:disstate}. 
After reviewing the forward Euler finite-difference scheme in 
\secref{sec:finite_difference}, we present the discretization/linearization scheme using variational equation in \secref{sec:variational_equation}. 
The resultant implementation of ILQR using variational equation that works as an MPC controller is presented and its advantage in online trajectory optimization is discussed in \secref{sec:ILQR_MPC}. 


\subsection{Finite-Difference Method} \label{sec:finite_difference}

  While ILQR requires linearized representations of the discrete-time dynamics, $f_x$ and $f_u$, shown in \eqref{eq:expansions}, no analytical expression is usually available for the state transition map $f$ of the discrete-time dynamics \eqref{eq:disstate}. 
  Thus, some numerical discretization methods are used. 
  In most of the existing studies using ILQR, the forward Euler finite-difference scheme is used, and the discrete-time system \eqref{eq:disstate} is approximated by 
  \begin{align} 
    x_{k+1} \approx x_k + \Delta t f_{\mathrm{c}}(x_k, u_k). \label{eq:finite_diff}
  \end{align}
  With this approximation, the linearizations of the transition map $f$ can be given as follows: for $i = 1, \dots, N $, 
  \begin{subequations} \label{eq:forward_euler}
  \begin{align}
       f_x(i) & = \dfrac{\partial f}{\partial x}\bigg{|}_{x=x_i} \approx  I_n + \Delta t \dfrac{\partial f_{\mathrm{c}} }{\partial x}\bigg{|}_{x=x_i}, \\
       f_u(i) & = \frac{\partial f}{\partial u}\bigg{|}_{u=u_i} \approx  \Delta t\frac{\partial f_{\mathrm{c}} }{\partial u}\bigg{|}_{u=u_i}.
  \end{align}
  \end{subequations}
  %
  The main drawback of the finite difference methods is the approximation accuracy deteriorates as the increase in the sampling period $\Delta t$.


\subsection{Variational Equation-Based Discretization} \label{sec:variational_equation}

  While the use of finite difference methods attempts to derive approximations of $f_x$ and $f_u$ through two  consecutive steps of discretization and linearization, the variational equation \cite{Hirsch2013} enables direct linearization of the transition map of the discretized dynamics. 
  Although variational equation is usually applied to nonlinear dynamical systems without control input, here we show that it can be applied to the nonlinear control system \eqref{eq:continuous} with zero-order holding of the control input. 
  To this end, consider the following augmented system defined for each time interval of $t \in [t_i, \,t_{i+1}]$ for $i = 0, \dots, N-1$: 
  \begin{equation}
    \dot{\tilde{x}} = \tilde{f}_{\mathrm{c}}( \tilde{x}) \label{eq:augmented_system}
  \end{equation}
  with  
  \begin{align}
    \tilde{x}:= \begin{bmatrix} x \\ u \end{bmatrix}, 
    \quad 
    \tilde{f}_{\mathrm{c}}(\tilde{x}) := 
    \begin{bmatrix} f_{\mathrm{c}}(x,u)  \\  0_m  \end{bmatrix},
  \end{align}
  and the initial condition $\tilde{x}_i := \tilde{x}(t_i) = [x_i^\top,\, u_i^\top]^\top$, where $0_m$ stands for the $m$-dimensional all-zero vector. 
  Let $\phi(t, \tilde{x}_i)$ be the solution of \eqref{eq:augmented_system} starting from the initial condition $\tilde{x}_i$ at time $t_i$, i.e., 
  \begin{align}
    \tilde x(t) =  \phi(t, \tilde{x}_i), \quad \forall t \in [t_i, \, t_{i+1}]. 
  \end{align}
  Since $\phi(t, \tilde{x}_i)$ is the solution of \eqref{eq:augmented_system}, we have the following equations:
  \begin{subequations}     
    \begin{align}
    \dot{\phi}(t, \tilde{x}_i) & = \tilde{f}_{\mathrm{c}}(\phi(t, \tilde{x}_i) ), \quad \forall t \in [t_i, \, t_{i+1}], \\
    \phi( t_i, \tilde{x}_i ) & = \tilde{x}_i. \label{eq:initial_phi}
    \end{align}
   \end{subequations}
  Differentiating these by the initial condition $\tilde{x}_i$, we obtain
  \begin{subequations}
   \begin{align}
    & \dfrac{\partial \dot{\phi}(t, \, \tilde{x}_i)}{\partial{\tilde{x}_i} }  =\frac{\partial \tilde{f}_{\mathrm{c}} (\tilde{x})}{\partial \tilde{x} }\bigg{|}_{\tilde{x}=\phi(t, \tilde{x}_i)}\dfrac{\partial \phi(t, \tilde{x}_i) }{\partial{\tilde{x}_i}},  
     \quad \forall t \in [t_i, \, t_{i+1}], \label{eq:phi}
    \\
    & \frac{\partial \phi (t_i, \, \tilde{x}_i)}{\partial{\tilde{x}_i}}  = I_n. \label{eq:I}
    \end{align}
  \end{subequations}
  Let $\Phi(t, \tilde{x}_i) = \partial \phi (t, \tilde{x}_i)/ \partial\tilde{x}_i $, then Eqs.\,\eqref{eq:phi} and \eqref{eq:I} can be rewritten as follows:
  \begin{subequations} \label{eq:variational_equation}
  \begin{align}
    & \dot{\Phi}(t, \tilde{x}_i) = \dfrac{\partial\tilde{f}_{\mathrm{c}}}{\partial \tilde{x} }\bigg{|}_{\tilde{x}=\phi(t,\tilde{x}_i)}\Phi(t, \tilde{x}_i), \quad \forall t \in [t_i, \, t_{i+1}],  \label{eq:Phi}  \\
    & \Phi(t_i, \tilde{x}_i) = I. \label{eq:initial_Phi}
  \end{align}
  \end{subequations}
  The above equation is the variational equation for the system \eqref{eq:augmented_system}, which is a matrix differential equation of $\Phi(t, \tilde{x}_i)$. 
  Since $\phi(t_{i+1}, \tilde{x}_i)$ gives $\tilde{x}_{i+1} = \tilde{x}(t_{i+1})$ and corresponds to the state transition map that maps $[x_i^\top, \, u_i^\top]^\top$ to the next state $[x_{i+1}^\top, \, u_i^\top]$,  the matrix $\Phi(t_{i+1}, \tilde{x}_i)$ gives the derivative of the transition map, i.e.,  
  \begin{equation} \label{eq:VE_based_linearization}
    \Phi(t_{i+1}, \tilde{x}_i)=
    \begin{bmatrix}
      \dfrac{\partial f(x, u)}{\partial x}\bigg{|}_{x=x_i, u=u_i} & \dfrac{\partial f(x, u)}{\partial u}\bigg{|}_{x=x_i, u=u_i}
      \\[1mm] 0_{m\times n} & I_m
    \end{bmatrix}, 
  \end{equation}
where $0_{m\times n}$ stands for the all-zero matrix of size $m\times n$. 
Therefore, by numerically integrating the variational equation \eqref{eq:variational_equation} together with \eqref{eq:augmented_system}, we obtain the linearizations $f_x(i)$ and $f_u(i)$. 
Note that the timestep of numerical integration of the variational equation can be chosen independently of the timestep $\Delta t$ of the discretization, and an accurate numerical solution can be obtain by using a standard ordinary differential equation solver.

\subsection{ILQR-MPC Implementation} \label{sec:ILQR_MPC}

For real-time feedback control of nonlinear systems, ILQR can be used in the framework of MPC. 
A standard ILQR-MPC implementation is shown in Algorithm\,\ref{alg:ilqr_MPC} with a simple setting of complete state observation and without consideration of the computational delay. 
Note that a slight modification is need in the real-time  implementation to take into account the delay for computing the control input $u_k$ after the current state $x_k$ is observed \cite{maciejowski2002}. 
In the algorithm, the main loop starting from the line 3 corresponds to real-time iteration $k=0, \dots K_{\mathrm{sim}}$, and at each time $k$,  the ILQR procedures are iterated for $N_{\mathrm{iter}}$ times. 
Since ILQR iteratively improves the estimate of the extremal trajectory, $N_{\mathrm{iter}}$ is an important hyper-paramater that affects the control performance.

The primal advantage of using variational equation is that the linearized model can be accurately computed independently of the timestep $\Delta t$ of the discretization, and this can result in a better control performance not only due to accurate linearization but also due to increase in the parameter $N_{\mathrm{iter}}$ as explained below. 
When the finite-difference methods are used, the timestep $\Delta t$ sometimes becomes too small, since it has to be chosen sufficiently smaller than the time scale of the fastest component of the system.  
Variational equation can be used to relax this stringent requirement. 
By enlarging the timestep $\Delta t$, more time can be available for ILQR trajectory optimization procedure at each time $k$. 
Furthermore, the number of horizons, $N$, becomes smaller when the prediction time interval is fixed. 
Thus, doubling the sampling period halves the size of the problem and doubles the time available for computation, and, as a result, the number of iteration $N_{\mathrm{iter}}$ scales with the square of the timestep $\Delta t$. 
The improvement of the control performance due to the above will be demonstrated in the numerical simulation provided below.


\begin{figure}[!t]
  \begin{algorithm}[H]
    \caption{MPC Using ILQR} \label{alg:ilqr_MPC}
    \begin{algorithmic}[1] 
    \STATE Initialize $x_0$ and $\bar{U} = [\bar{U}(0), \dots \bar{U}(N-1)]^\top$ 
    \STATE $\bar{X} \leftarrow $ Simulate from $x_0$ using $\bar{U}$ and \eqref{eq:disstate}
    \FOR {$k = 0: 1 : K_{\mathrm{sim}}$}
        \STATE Obtain current state $x_k$ of controlled system
        \STATE /* ILQR trajectory optimization */
        \FOR{ $j = 1: 1: N_{\mathrm{iter}} $}
        \STATE /* Linearization */
        \FOR {$i = k:1:k+N$}
        \STATE Calculate linearized model $f_x(i)$, $f_u(i)$ using \eqref{eq:forward_euler} (finite-difference method) or \eqref{eq:VE_based_linearization} (variational equation) and calculate cost derivatives $l_x(i)$, $l_u(i)$, $l_{xx}(i)$, $l_{uu}(i)$, $l_{ux}(i)$ 
      \ENDFOR
      \STATE /* Backward pass */
      \STATE Calculate $V_x(k+N)$,$V_{xx}(k+N)$ by \eqref{eq:boundary condition}
      \FOR {$i = k+N-1: -1 : k$}
        \STATE Calculate coefficients $Q_x(i)$, $Q_u(i)$, $Q_{xx}(i)$, $Q_{uu}(i)$, and $Q_{ux}(i)$ using \eqref{eq:expansions} and \eqref{eq:value_derivatives} 
        \STATE Calculate $d(i)$ and $K(i)$ using \eqref{eq:opt_control}
      \ENDFOR
      \STATE /* Forward pass */ 
      \FOR {$i = 0:1:N-1$}
        \STATE Calculate $\hat{X}$ and $\hat{U}$ using \eqref{eq:forward_pass} for each $\alpha$ 
      \ENDFOR
      \STATE $\bar{X},\, \bar{U} \gets \hat{X}$, $\hat{U}$ that minimize the cost function
      \ENDFOR
      \STATE /* Recede horizon /*
      \STATE Apply first element of $\bar{U}$ as $u_k$
      \STATE $\bar{U} \gets [\bar{U}(1), \dots \bar{U}(N-2), \bar{U}(N-2)]^\top$ 
      \ENDFOR
      \end{algorithmic}
  \end{algorithm}
  \end{figure}

\section{Numerical Simulations} \label{sec:simulations}

In this section, we 
demonstrate the effectiveness of ILQR with variational-equation based discretization. 
The problem setting of the swing-up stabilization of an inverted pendulum is introduced in \secref{sec:inverted_pendulum}.
We then show numerical simulations of trajectory optimization as an open-loop optimal control problem in \secref{sec:trajectory_optimization}. 
The simulation results for ILQR-based model predictive controller are presented in \secref{sec:ILQR-MPC_simulation}.   
All the 
simulations were performed using a desktop computer 
with Intel\textregistered Core\texttrademark i9-10900X CPU @3.70GHz and with 48 GB RAM, 
and the codes were implemented in MATLAB R2023a. 

\subsection{Swing-up Control of Inverted Pendulum} \label{sec:inverted_pendulum}
\begin{figure}[!t]
    \centering
    \includegraphics[width=0.4\linewidth]{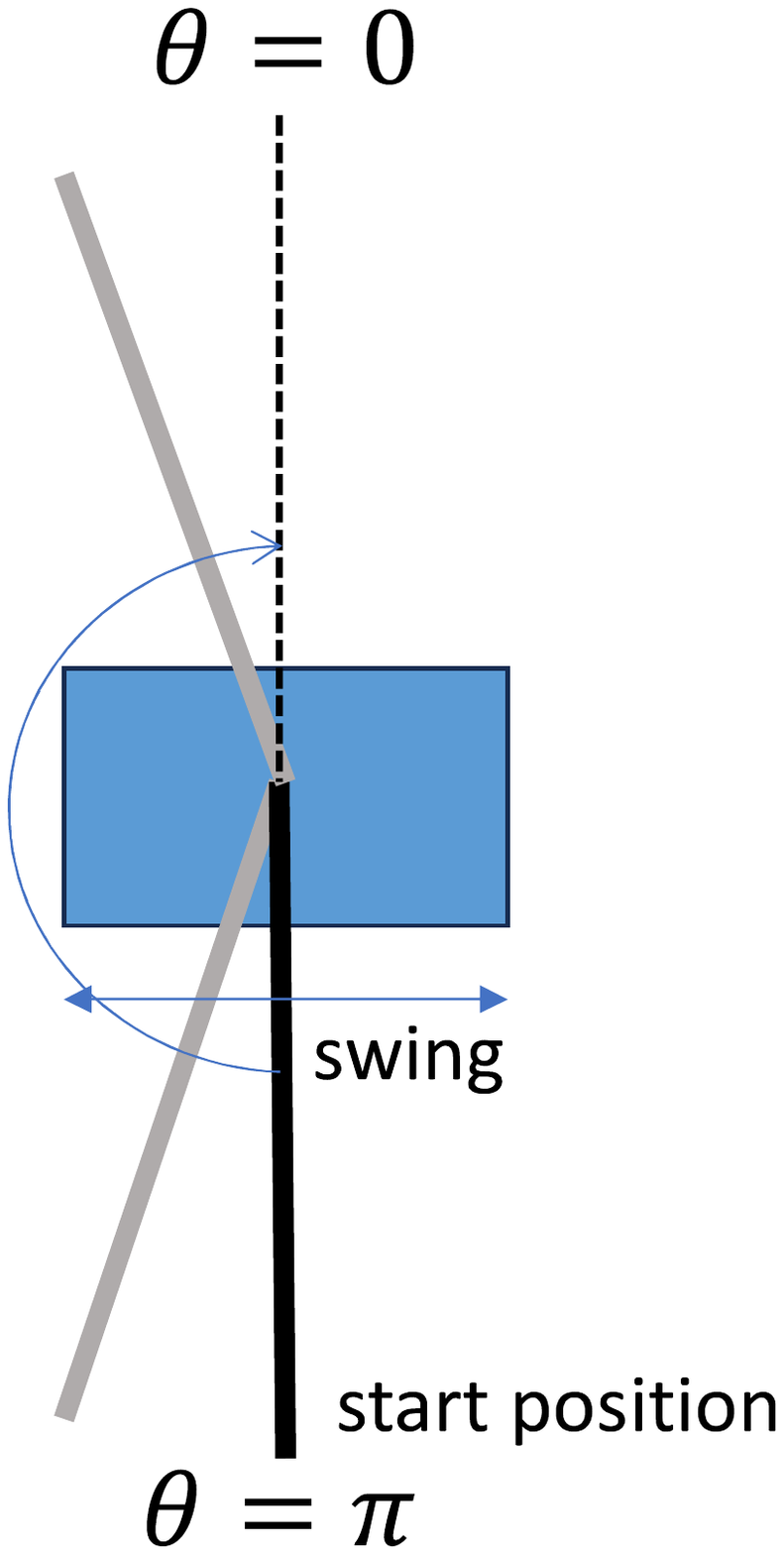}
    \caption{The inverted pendulum model}
    \label{fig:pendulum}
\end{figure}

The inverted pendulum system considered here is a two dimensional model studied in \cite{Sakamoto2013}, where the pendulum is attached on a mass-less cart (\figref{fig:pendulum}). 
The equation of motion of this inverted pendulum model is given by the following input-affine form: 
\begin{align}
  \dot{x} = 
  f_{\mathrm{p}}(x) +g_{\mathrm{p}}(x) u   
  \end{align}
  where $ x:=[x_1, x_2]^\top = [\theta, \dot{\theta}] \in \mathbb{R}^2$ and $u \in \mathbb{R}$. Here $x_1 = \theta $ stands for the angle of the pendulum, $x_2 = \dot{\theta}$ for the angular velocity, and $u$ for the acceleration of the cart. The functions $f_{\mathrm{p}}$ and $g_{\mathrm{p}}$ are given by
\begin{subequations}
\begin{align}
  f_{\mathrm{p}}(x) & =
  \begin{pmatrix}
  x_2 \\[2pt]
  \dfrac{MGL\sin{x_1}-ML^2{x_2}^2\sin{x_1}\cos{x_1}}{J+ML^2\sin^2{x_1}}
  \end{pmatrix}, \\
  g_{\mathrm{p}}(x) & = 
  \begin{pmatrix}
  0 \\[2pt]
  \dfrac{-L\cos{x_1}}{J+ML^2\sin^2{x_1}}
  \end{pmatrix}. 
\end{align}
\end{subequations}%
where $M$ stands for the mass of the pendulum, $G$ for the acceleration of gravity, $L$ for the length of the pendulum to the center of gravity, and $J$ for the moment of inertia. 
The setting of the parameter is the same as in \cite{Sakamoto2013}.

The swing-up control can be formulated as a standard optimal control problem with the following quadratic objective function:
\begin{align}
 J_{\mathrm{c}} = \int_0^\infty x^\top Q_{\mathrm{s}} x + u^\top R u \mathrm{d}t 
 \label{eq:cost_continuous}
\end{align}
with $Q_{\mathrm{s}} = \mathrm{diag}\{ [2,\,0.01]^\top \}$ and $R =2$. 
Note that the origin $x = [0,\,0]^\top$ corresponds to the upright position, and we consider a swing-up task from the pending position of $x = [-\pi,\,0]^\top$. 
The optimal state and input trajectories can be obtained by the stable manifold method \cite{Sakamoto2008,Sakamoto2013} for numerically solving the Hamilton–Jacobi-Bellman equation. 
The numerical algorithm is reviewed in Appendix~\ref{sec:stable_manifold_method}.

\begin{figure}[t!]
  \begin{subfigure}[t]{0.5\textwidth} 
		\vbox{
		\centering{
		  \includegraphics[width=0.95\linewidth]{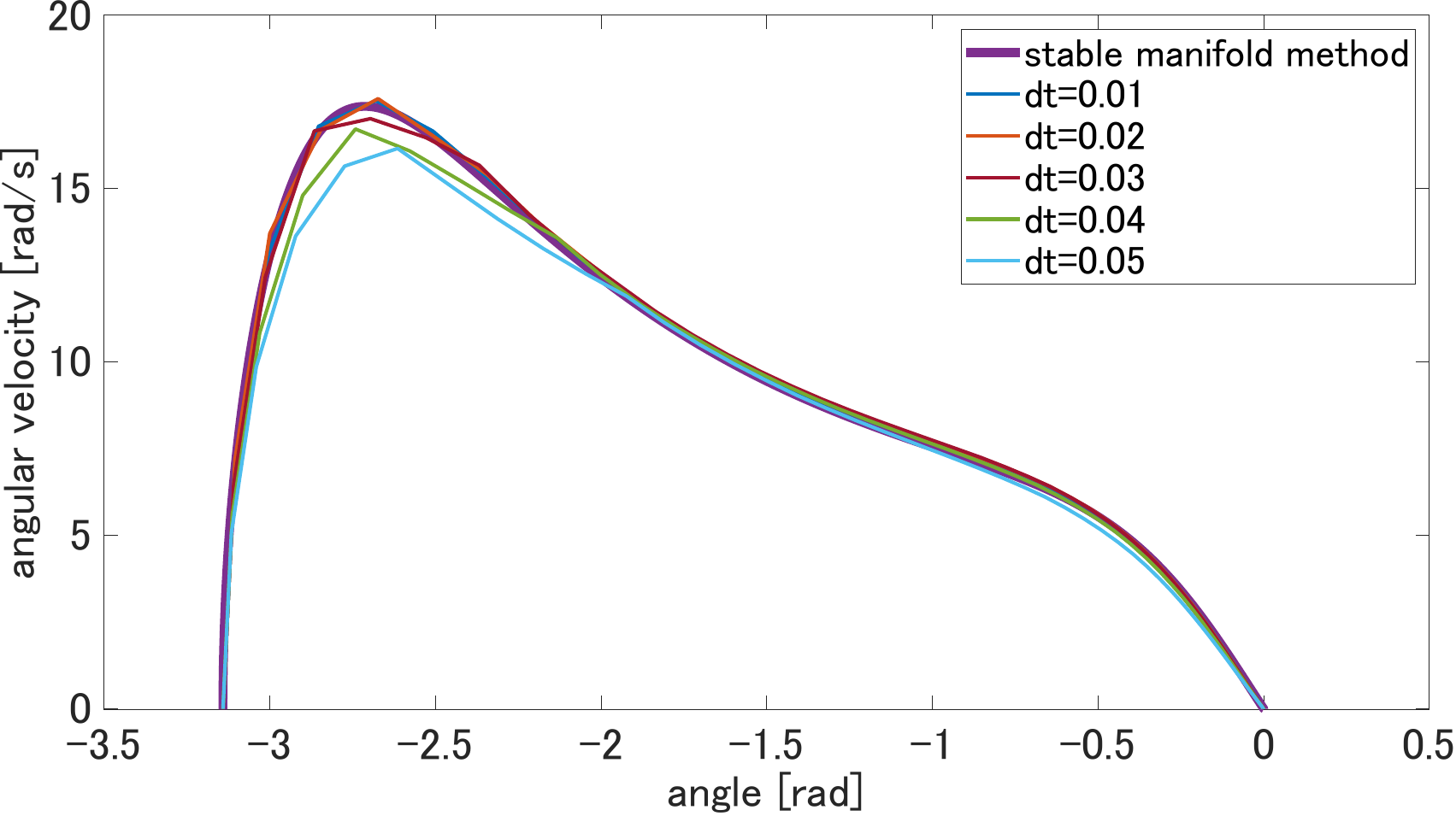}
		}%
		}%
		\subcaption{Finite-difference method} \label{fig:Trajectory_finite difference}
		\end{subfigure}%
		\begin{subfigure}[t]{0.5\textwidth} 
		\centering{
		\centering\includegraphics[width=0.95\linewidth]{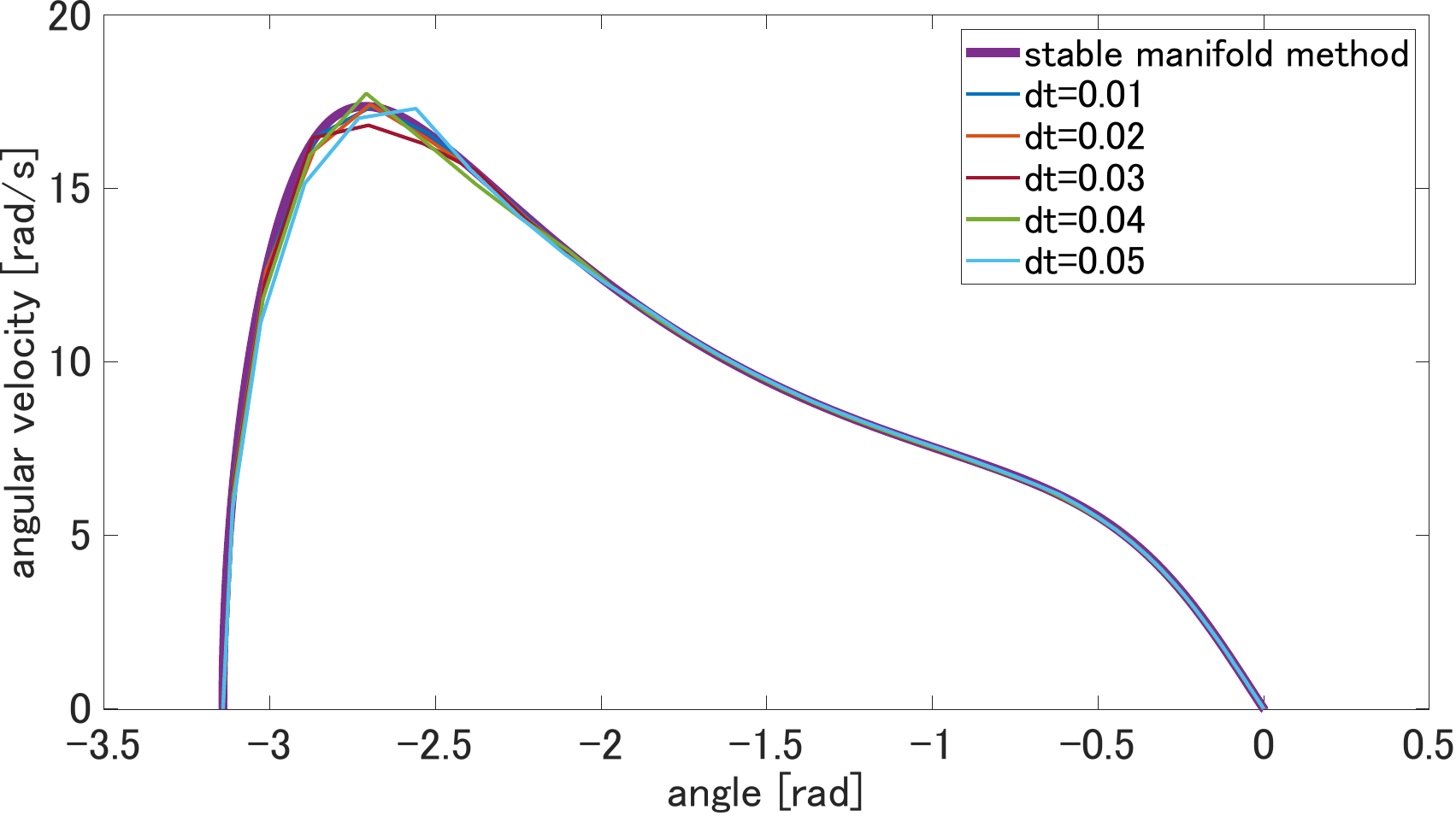}
		\subcaption{Variational equation} \label{fig:Trajectory_variational equation}
		}\end{subfigure}%
		\caption{Comparison of state trajectories for various settings of timestep $\Delta t$ from $\SI{0.01}{s}$ to $\SI{0.05}{s}$. }  \label{fig:Trajectory}
	\end{figure}

\subsection{Trajectory Optimization} \label{sec:trajectory_optimization}

%
In this subsection, for the purpose of clarifying the benefits of using variational equation in terms of approximation accuracy, we present numerical simulations of trajectory optimization solved as an open-loop optimal control problem. 
The cost integral \eqref{eq:cost_continuous} of the original continuous-time problem was translated to the cost function of a discrete-time problem by using the integrand for the cost term, and the functions $l$ and $l_{\mathrm{f}}$ in \eqref{eq:cost_discrete} are given as 
\begin{subequations}
\begin{align} 
    l(x,\,u) & =   x^\top Q_{\mathrm{s}} x + u^\top R u,  \\
    l_{\mathrm{f}}(x) & =   x^\top Q_{\mathrm{f}} x 
\end{align}
\end{subequations}
where we chose $Q_{\mathrm{f}} = Q_{\mathrm{s}}$. 
The initial condition is fixed at the pending position: $x_0 = [-\pi,\, 0]^\top$. 
The horizon $N$ is chosen for each timestep $\Delta t$ such that the total time $T=N \Delta t$ becomes (approximately) equal to $ \SI{0.8}{s}$. 
For example, $N = 80$ for $\Delta t = \SI{0.01}{s}$, and $N = 16$ for $\Delta t = \SI{0.05}{s}$. 
The initial control sequence is set to a series of all zeros, and  the procedure of ILQR is iterated for $N_{\mathrm{iter}}= 200$ times, such that the ILQR iteration converges sufficiently. 

\Figref{fig:Trajectory} shows the resultant state trajectories obtaind by ILQR using (a) finite-difference (forward Euler) method and (b) variational equation under several settings of $\Delta t = \SI{0.01}{s}$ to $\SI{0.05}{s}$. 
As shown in \figref{fig:Trajectory_finite difference}, the trajectory with
finite-difference method becomes far from the optimal trajectory calculated by the stable manifold method (shown by the purple line) when the timestep $\Delta t$ is larger than $\SI{0.03}{s}$. 
This is clearly due to a large approximation error in the linearized model ($f_x(i)$ and $f_u(i)$) calculated based on finite-difference approximations.  
In contrast, when the variational equation is used, as shown in \figref{fig:Trajectory_variational equation}, there is no pronounced effect on the resultant state trajectory by enlarging the timestep $\Delta t$ up to $\SI{0.05}{s}$, showing the effectiveness of the accurate calculation of the linearized discrete-time model. 
   
\begin{figure}[!t]
    \centering\includegraphics[width=0.6\linewidth]{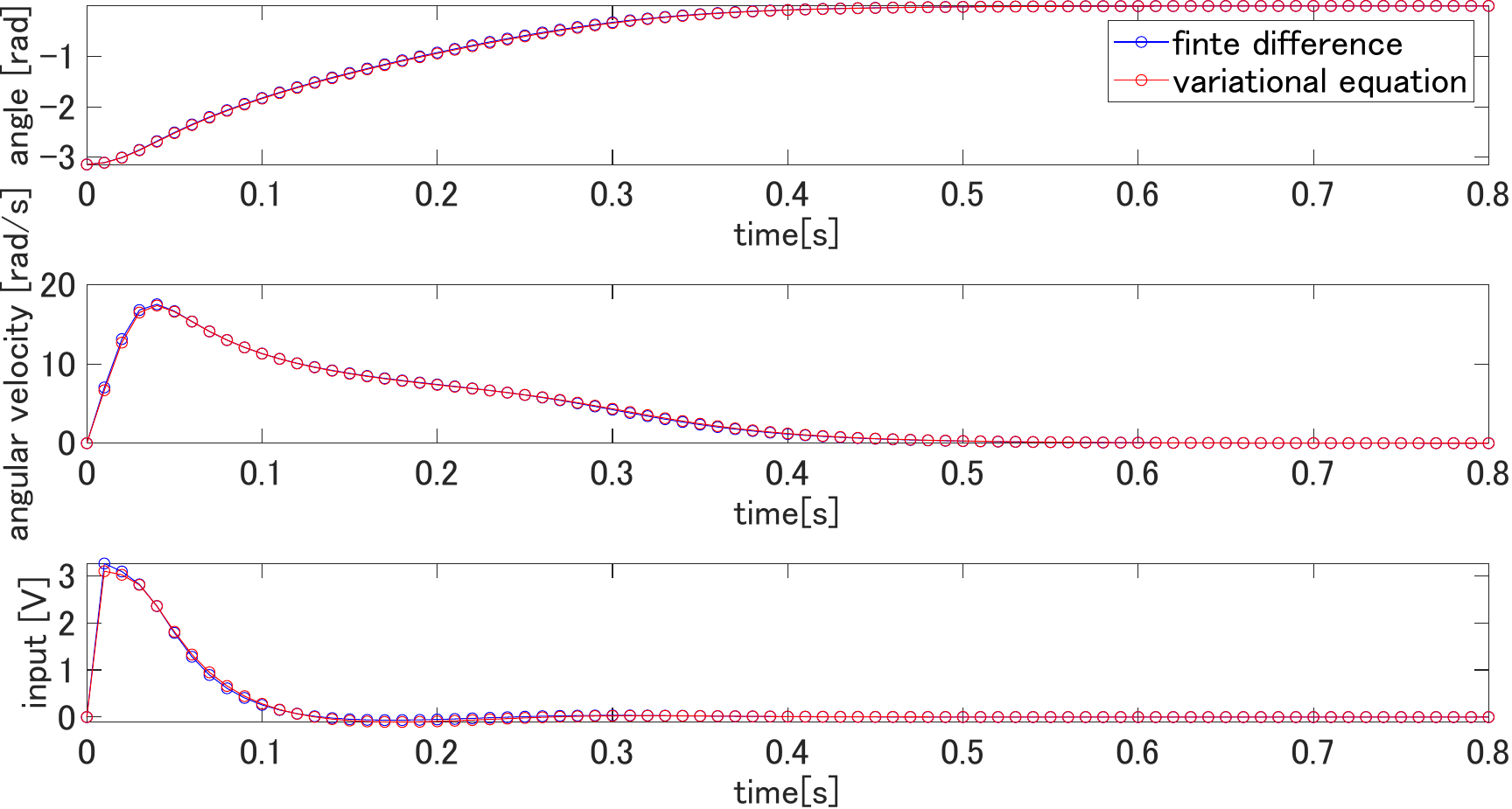}
    \caption{Comparison of time courses of the state and input for the timestep $\Delta t = \SI{0.01}{s}$.}
    \label{fig:time_variational}
\end{figure}

\begin{figure}[!t]    
    \centering
    \includegraphics[width=0.6\linewidth]{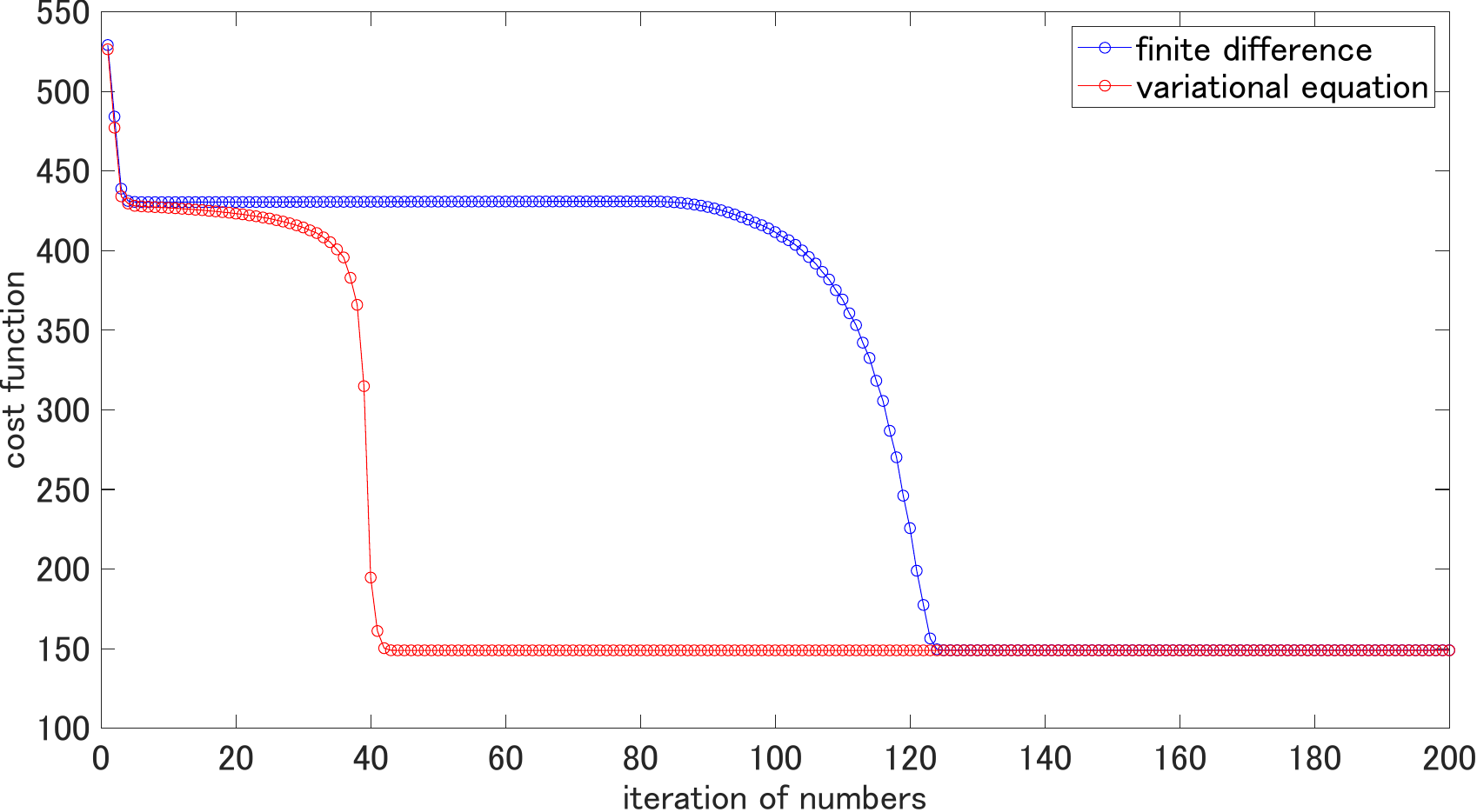}
    \caption{Cost vs. iterations for the timeste $\Delta t = \SI{0.01}{s}$.}
    \label{fig:cost}
\end{figure}

Next, we further discuss the results for the timestep $\Delta t = \SI{0.01}{s}$, including the iteration processes of the trajectory optimization. 
\Figref{fig:time_variational} shows the time courses of the state and input after the convergence of the ILQR iteration, 
    and it can be confirmed that there is no apparent difference in the converged trajectories. 
\Figref{fig:cost} shows the value of the cost function at each iteration, and 
the ILQR algorithm using variational equation converges faster than that using finite-difference method. 
That is, the use of variational equation has the advantage not only for computing accurate trajectories for a large $\Delta t$, but also for faster convergence of ILQR iterations.

\subsection{Model Predictive Controller} \label{sec:ILQR-MPC_simulation}

\begin{table*}[!t]
 \caption{Error between closed-loop and optimal trajectories for ILQR with finite-difference method.}
 \label{table:error_difference}
  \centering
  \fontsize{8}{10}\selectfont
  \begin{tabular}{lllllllllllll}
    \hline \diagbox{$\Delta t$}{Iteration} & 1 & 2 & 3 & 4 & 5 & 6 & 7 & 8 \\ \hline
    0.01 & 9.87E-02 & 2.21E-02 & 8.02E-03 & 7.60E-04 & 5.59E-04 & 7.36E-04 & 7.70E-04 & 7.86E-04 \\ \hline
    0.02 & \cellcolor[rgb]{0.9, 0.9, 0}2.74E-01 & \cellcolor[rgb]{0.9, 0.9, 0}1.07E-01 & 4.56E-02 & 1.78E-02 & 9.82E-04 & 2.22E-03 & 3.45E-03 & 3.50E-03 \\ \hline
    0.03 & \cellcolor[rgb]{0.9, 0.9, 0}4.27E-01 & \cellcolor[rgb]{0.9, 0.9, 0}2.35E-01 & \cellcolor[rgb]{0.9, 0.9, 0}1.08E-01 & \cellcolor[rgb]{0.9, 0.9, 0}3.99E-02 & 2.59E-03 & 1.79E-03 & 1.79E-03 & 1.85E-03 \\ \hline
    \rowcolor[rgb]{0.9, 0.9, 0} \cellcolor{white}0.04 & 6.72E-01 & 5.29E-01 & 2.83E-01 & 1.18E-01 & 5.18E-02 & 1.24E-02 & 1.23E-02 & 1.75E-02 \\ \hline
    \rowcolor[rgb]{0.9, 0.9, 0} \cellcolor{white}0.05 & 8.95E-01 & 7.97E-01 & 4.72E-01 & 2.22E-01 & 1.11E-01 & 3.35E-02 & 3.00E-02 & 4.26E-02 \\ \hline 
  \end{tabular}\\[1mm]
  {\footnotesize The letter E is used for the scientific notation, that is, $m \mathrm{E} n$ represents $m \times 10^n$. \hspace{60mm} } 
  \end{table*}
\begin{table*}[!t]
  \caption{Error between closed-loop and optimal trajectories for ILQR wiht variational equation. }
  \label{table:error_variational}
    \centering
    \fontsize{8}{10}\selectfont
    \begin{tabular}{lllllllllllll}
    \hline \diagbox{$\Delta t$}{Iteration} & 1 & 2 & 3 & 4 & 5 & 6 & 7 & 8 \\ \hline
    0.01 & 6.59E-02 & 1.82E-02 & 1.67E-03 & 3.99E-04 & 4.96E-05 & 1.88E-05 & 1.34E-05 & 1.22E-05 \\ \hline
    0.02 & \cellcolor[rgb]{0.9, 0.9, 0}1.76E-01 & \cellcolor[rgb]{0.9, 0.9, 0}5.69E-02 & 6.59E-03 & 2.06E-03 & 7.45E-04 & 5.50E-04 & 5.02E-04 & 4.89E-04 \\ \hline
    0.03 & \cellcolor[rgb]{0.9, 0.9, 0}2.79E-01 & \cellcolor[rgb]{0.9, 0.9, 0}8.86E-02 & \cellcolor[rgb]{0.9, 0.9, 0}1.25E-02 & \cellcolor[rgb]{0.9, 0.9, 0}4.32E-03 & 2.21E-03 & 1.79E-03 & 1.68E-03 & 1.64E-03 \\ \hline
    \rowcolor[rgb]{0.9, 0.9, 0} \cellcolor{white}0.04 & 4.34E-01 & 1.36E-01 & 1.44E-02 & 6.40E-03 & 4.01E-03 & 3.60E-03 & 3.50E-03 & 3.47E-03 \\ \hline
     \rowcolor[rgb]{0.9, 0.9, 0} \cellcolor{white}0.05 & 6.37E-01 & 1.59E-01 & 1.80E-02 & 1.47E-02 & 1.27E-02 & 1.25E-02 & 1.25E-02 & 1.25E-02 \\ \hline
    \end{tabular}
    \end{table*}

Here we discuss benefits of the proposed method in terms of the real-time computation. 
Feedback stabilization of inverted pendulum is simulated with a model predictive controller with Algorithm\,\ref{alg:ilqr_MPC}. 
The prediction interval is set to $T=\SI{0.4}{s}$, by considering that the pendulum can be swung up nearly in this interval as shown in \figref{fig:time_variational}. 

\begin{figure}[!t] 
    \centering{
    \includegraphics[width=8.3cm]{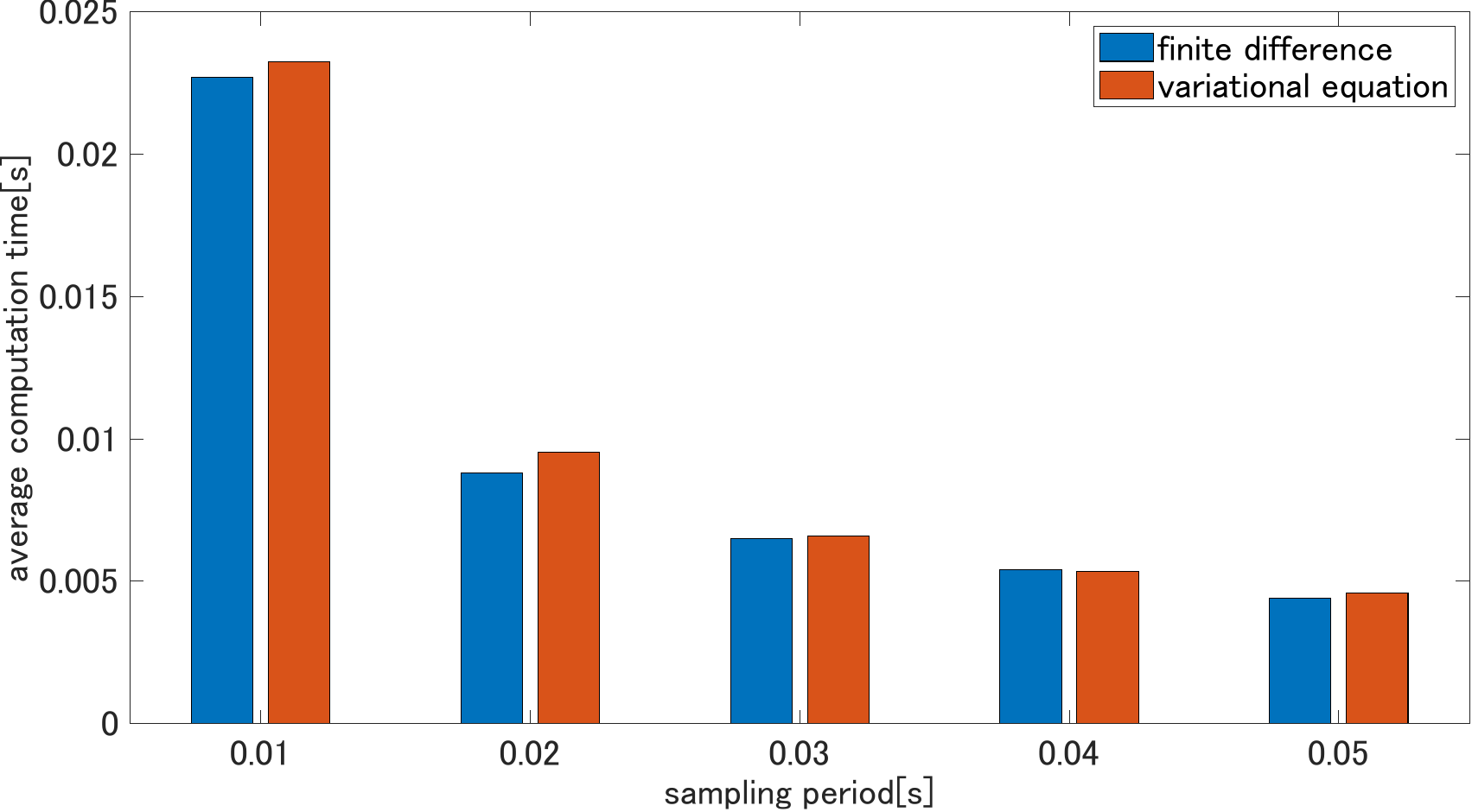}
    \caption{Averaged computational time for a single iteration of ILQR.}
    \label{fig:bar_solvetime}
    }\end{figure}%

Tables \ref{table:error_difference} and \ref{table:error_variational} show the mean squared error between the closed-loop and optimal trajectories for each setting of the timestep $\Delta t$ and the number of iteration $N_{\mathrm{iter}}$ at each time.  
As an overall trend, it can be seen that the error tends to decrease as the number of iteration increases, but the error increases as the increase in $\Delta t$.  
In these tables, the cells highlighted in yellow represent the conditions of $\Delta t$ and $N_{\mathrm{iter}}$ that could be executable by real-time computation.  
This is determined by the averaged computational time for a single iteration of ILQR (from line 7 to 21 in Algorithm\,\ref{alg:ilqr_MPC})  shown in \figref{fig:bar_solvetime}.  As the increase in the timestep $\Delta t$, the computational time is reduced due to the reduction in the computational burden especially in the backward pass. 
It is notable that the increase in the computational time for using variational equation (solving \eqref{eq:variational_equation}) is small compared with the entire procedure of the single iteration.  
As shown in Tables \ref{table:error_difference} and \ref{table:error_variational}, while real-time computation with $\Delta t = \SI{0.01}{s}$ is infeasible with the given hardware, it becomes feasible for $\Delta t \ge \SI{0.02}{s}$ and the number of iteration can be significantly increased by increasing the timestep $\Delta t$.  
    The highlighted cells in the tables will be shifted to the right 
    when using a more powerful hardware, and vice versa. 

\begin{figure}[!t]
    \begin{subfigure}[t]{0.5\textwidth} 
    \vbox{	
    \centering
      \includegraphics[width=0.95\linewidth]{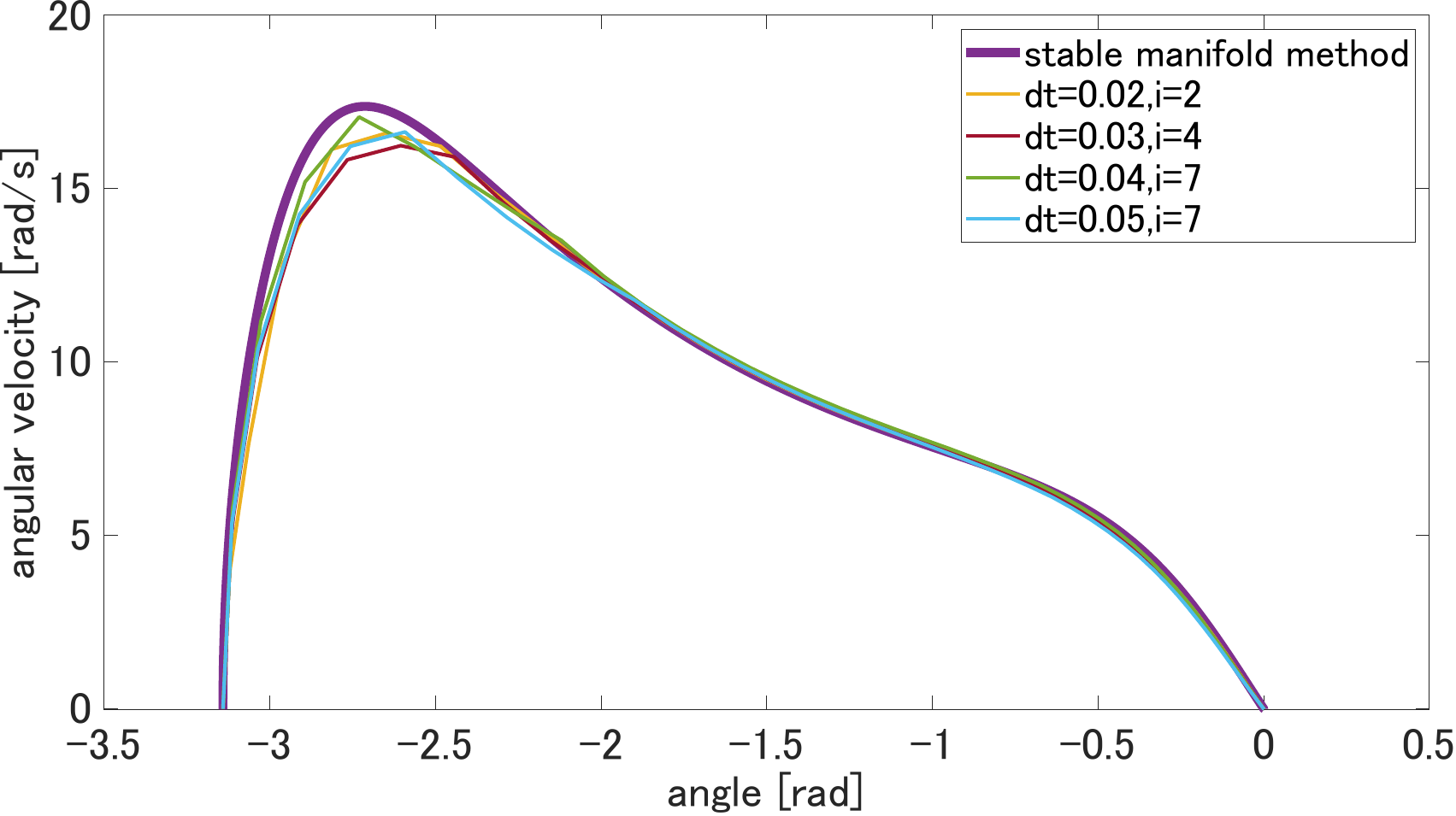}
    }%
    \subcaption{Finite-difference method}
    \label{fig:MPC_difference}
    \end{subfigure}%
    \begin{subfigure}[t]{0.5\textwidth} 
    \centering{
    \includegraphics[width=0.95\linewidth]{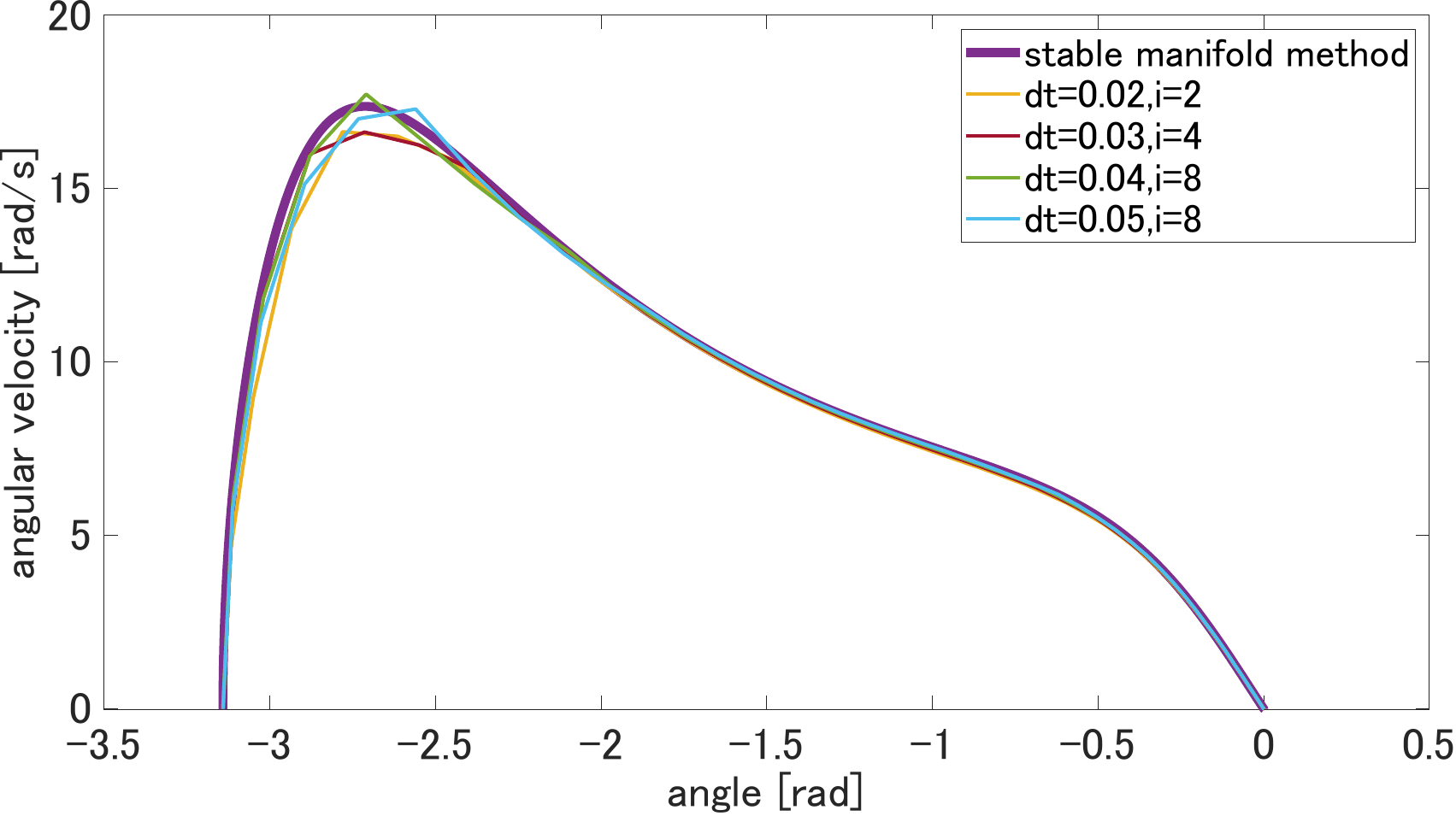}
    \subcaption{Variational equation} \label{fig:MPC_variational}
    }\end{subfigure}%
    \caption{Closed-loop trajectories with ILQR-MPC } \label{fig:MPC}
\end{figure}

\begin{figure}[!t]
    \centering
      \includegraphics[width=8.3cm]{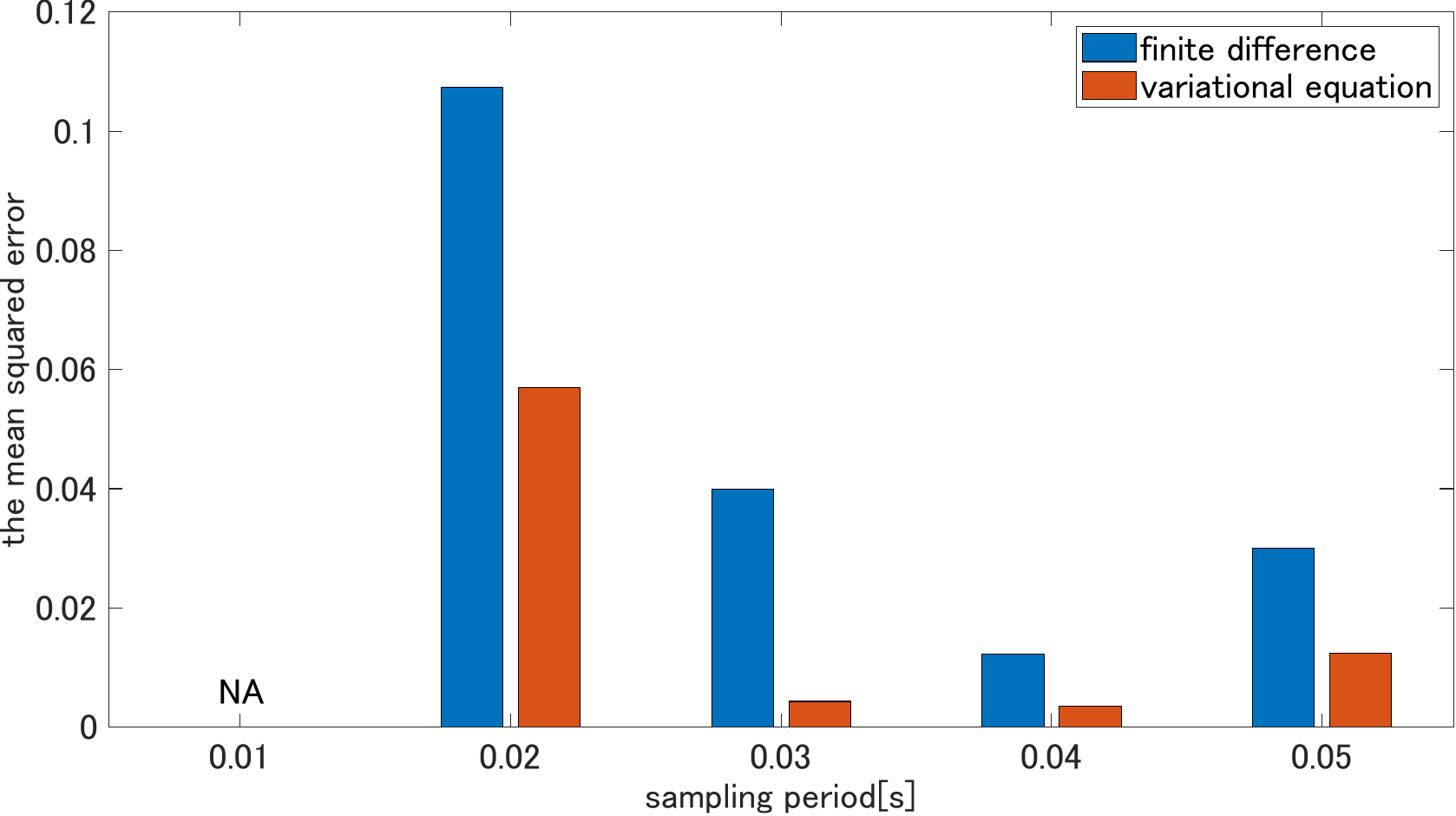}
    \caption{Error from optimal trajectory achieved by real-time computation. }
    \label{fig:bar_error}
\end{figure}

\Figref{fig:MPC} shows the best trajectory that is executable in real-time for each setting of the timestep $\Delta t$. 
The mean squared error from the optimal trajectory is summarized in \figref{fig:bar_error}.  
%
In the current setting, the error decreases as the increase in $\Delta t$ up to $\SI{0.04}{s}$ in the both cases of using finite-difference method and variational equation. 
This is mainly due to the increase in the number of iteration $N_{\mathrm{iter}}$. 
However, the errors are much larger when the finite-difference method is used due to the low accuracy of the discretization. 
The difference between the errors for the two methods is most pronounced at $\Delta t= \SI{0.03}{s}$. 
At $\Delta t= \SI{0.04}{s}$, although the error with the finite-difference method is decreased due to the increase in the iteration ($N_{\mathrm{iter}} = 7$), this is nearly three point five times that with the variational equation with the same condition and is worse than the variational equation with  $N_{\mathrm{iter}} = 4$ (See Table\,\ref{table:error_variational}). 
At $\Delta t= \SI{0.05}{s}$, the error rises for both of the methods.  
This increase in the errors is caused by the lower update frequency of the control input.
In summary, ILQR with variational equation achieved the best control performance at $\Delta t= \SI{0.04}{s}$ thanks to both of the increase in the number of iteration and the accurate calculation of the linearized model.


\section{Conclusions} \label{sec:conclusions}

This paper discussed discretization methods for the implementation of Iterative Linear Quadratic Regulator (ILQR). 
While finite-difference approximations have usually been used to derive a discrete-time state equation from the original continuous-time model, we proposed to use variational equation to directly calculate linearizations of the state transition map of the discretized system. 
By using variational equation, it is possible to increase the timestep with a high discretization accuracy, and this leads to a better control performance by increasing the possible number of ILQR iterations in real-time computation. 
The effectiveness of the proposed method was  demonstrated through numerical simulations of the swing-up control of an inverted pendulum, for which the exact optimal trajectory was available to quantify the control performance. 

Future work includes application of the proposed method in an experimental setting and extensions of the method to consider additional constraints through Augmented Lagrangian \cite{Howell2019} or hybrid dynamical systems \cite{Kong2023}.

\section*{Acknowledgments}

This work was supported by Grant-in-Aid for Scientific Research (KAKENHI) from the Japan Society for Promotion of Science (\#23K13354).

\appendix
\section{The stable manifold method} \label{sec:stable_manifold_method}

    This appendix briefly reviews the stable manifold method \cite{Sakamoto2008} to calculate the exact optimal trajectory for swing-up stabilization of an inverted pendulum. 
    Consider the control problem with the state equation and the cost function given by 
      \begin{gather}
        \dot x =f(x)+g(x)u,\quad x(0)=x_0 \label{eq:system}, \\
        J=\int_{0}^{\infty}(x^\top Qx+u^\top Ru)\mathrm{d}t. \label{eq:j}
      \end{gather}
    where $x\in \mathbb{R}^n$,  $u\in \mathbb{R}^m$, $f(\cdot):\mathbb{R}^n\rightarrow\mathbb{R}^n$, $g(\cdot):\mathbb{R}^n\rightarrow\mathbb{R}^{n\times m}$.
    We assume that $f(0)=0$ and write $f(x)$ as $f(x) = A x + \mathcal{O}(|x|^2)$ where $A$ is an $n \times n$ real matrix. 

    For this control problem, the Hamilton-Jacobi-Bellman equation is expressed as
    \begin{align}
      H(x,\lambda)=\lambda^Tf(x)-\frac{1}{4}g(x)R^{-1}g(x)^T\lambda+x^TQx=0. \label{eq:H}
    \end{align}
    where $\lambda_1 = \partial V/\partial x_1, \dots, \lambda_n = \partial V/\partial x_n$ with $V(x)$ an unknown
function. 
    The stabilizing solution of Eq. \eqref{eq:H} is equivalent to the stable manifold of the origin of the following Hamiltonian system: 
      \begin{gather}
      \left \{
      \begin{aligned}
      &\dot{x}=\frac{\partial{H}}{\partial{\lambda}}(x,\lambda),   \\
      &\dot{\lambda}=-\frac{\partial{H}}{\partial{x}}(x,\lambda).\label{eq:canonical_equation}
      \end{aligned}
      \right.  
      \end{gather}
      Then, by introducing a suitable coordinate transformation $T$ as 
      \begin{align}
        \begin{pmatrix}
          q \\
          p
          \end{pmatrix}
          := T^{-1}
          \begin{pmatrix}
            x \\
            \lambda
            \end{pmatrix},
      \end{align}
      the system \eqref{eq:canonical_equation} can be rewritten as 
      \begin{align}
        \begin{pmatrix}
        \dot{q}  \\
        \dot{p}
        \end{pmatrix}
        =
        \begin{pmatrix}
        A-\bar{R}\Gamma && 0  \\
        0 && -(A-\bar{R}\Gamma)^T
        \end{pmatrix}
        \begin{pmatrix}
        q   \\
        p 
        \end{pmatrix}
        + \text{higher order terms} \label{eq:transAfter}
        \end{align}
        where $\bar{R}=g(x)R^{-1}g(x)^\top$ and $\Gamma$ is the stabilized solution of the Riccati equation consisting of linear terms of the form \eqref{eq:H}.
    
        Furthermore, by setting $F=A-\bar{R}\Gamma$ and letting $n_s$ and $n_u$ be higher-order nonlinear terms, Eq. \eqref{eq:transAfter} can be rewritten  as follows:
        \begin{equation}
          \begin{cases}
          &\dot{q}={F}q+n_s(t,q,p), \\
          &\dot{p}=-{F}^\top p+n_u(t,q,p). 
          \end{cases} \label{eq:stable}
          \end{equation}
          We assume that $F$ is an asymptotically stable $n\times n$ real matrix, and $n_s$ and $n_u$ are higher order terms with sufficient smoothness.
          We define the sequences $q_k(t,\xi)$ and $p_k(t,\xi)$ by
          \begin{equation}
            \left \{
            \begin{aligned}
            q_{k+1}&=\mathrm{e}^{Ft}\xi+\int_{0}^{t} \mathrm{e}^{F(t-s)} n_s(s,q_k(s),p_k(s))ds, \\
            p_{k+1}&=-\int_{t}^{\infty}\mathrm{e}^{-F^\top(t-s)}n_u(s,q_k(s),p_k(s))ds,
            \end{aligned}
            \right. \label{eq;stable manihold}
            \end{equation}
            for $k = 0,\,1, \,2, \dots,$ and
            \begin{equation}
              \left \{
              \begin{aligned}
              q_{0}&=\mathrm{e}^{Ft}\xi, \\
              p_{0}&=0,
              \end{aligned}
              \right.
            \end{equation}
            with arbitary $\xi \in \mathrm{R}^n$. Then the following theorem holds:
        \begin{theorem}
          \cite{Sakamoto2008}. The sequences $q_k(t,\xi)$ and $p_k(t,\xi)$ are convergent to zero for sufficiently small $|\xi |$, that is, $q_k(t,\xi)$, $p_k(t,\xi) \rightarrow 0$ as $t \rightarrow \infty$ for all $k = 0, 1, 2, ...$.
          Furthermore, $q_k(t,\xi)$ and $p_k(t,\xi)$ are uniformly convergent to a solution of \eqref{eq:stable} on $[0, \infty)$ as $k \rightarrow \infty$ for sufficiently small $|\xi|$. Let $q(t,\xi)$ and $p(t,\xi)$ be the limits of $q_k(t,\xi)$ and $p_k(t,\xi)$, respectively. Then, $q(t,\xi)$, $p(t,\xi)$ are the solution on the stable manifold of \eqref{eq:stable}, that is, $q(t,\xi)$, $p(t,\xi) \rightarrow 0$ as $t \rightarrow \infty$.
        \end{theorem}
        For each $k$, the functions in \eqref{eq;stable manihold} are calculated for the system \eqref{eq:transAfter} to obtain $p_k(t,\xi)$ and $q_k(t,\xi)$. Functions $x_k(t,\xi)$ and $\lambda(t,\xi)$ given by 
        \begin{align}
          \begin{pmatrix}
            x \\
            \lambda
            \end{pmatrix}
            = T
            \begin{pmatrix}
              p \\
              q
              \end{pmatrix}
        \end{align}
        form parameterizations of the approximate stable manifold.
        Since the obtained trajectories are solutions of the Hamilton-Jacobi-Bellman equations, the computed trajectories are guaranteed to be optimal.

\bibliographystyle{ieeetr}  
\bibliography{ASME_ILQR}

\end{document}